\documentclass[12pt]{iopart}
\usepackage{graphicx,stackengine,cite}
\usepackage[colorlinks,linkcolor=blue,citecolor=blue]{hyperref}  \usepackage{url}
\usepackage{verbatim}
\usepackage{xcolor}

\newcommand{\R}{\mathbb{R}}

\newcommand{\vi}{\bi{v}}

\newcommand{\x}{\bi{x}}
\newcommand{\y}{\bi{y}}
\newcommand{\z}{\bi{z}}
\newcommand{\vr}{\bi{r}}

\newcommand{\q}{\bi{q}}

\newcommand{\Dq}{\Delta \bi{q}(t,t')}

\newcommand{\uveci}{{\hat{\biota}}}

\newcommand{\lt}{\left}
\newcommand{\rt}{\right}
\newcommand{\te}{\theta}

\newcommand{\eL}{{\cal L}}
\newcommand{\eps}{\bepsilon}

\newcommand{\cdt}{\!\cdot\!}
\newcommand{\tc}{t_{\rm c}}
\newcommand{\ta}{t_{\rm a}}

\newcommand{\thst}{\theta_{\rm st}}

\newcommand{\colrev}[1]{{\color{black} #1}} 

\usepackage{iopams}
\usepackage{setstack} 
\usepackage{tocloft} 
\usepackage{subcaption}

\begin{document}

\title[The multivalued solution of the diffusion equation]{
Dynamical theory of topological defects I: the multivalued solution of the diffusion equation}

\author{Jacopo Romano$^1$, Beno\^it Mahault$^{1,*}$, Ramin Golestanian$^{1,2,\dagger}$}

\address{$^1$Max Planck Institute for Dynamics and Self-Organization (MPI-DS), 37077 G\"ottingen, Germany}
\address{$^2$Rudolf Peierls Centre for Theoretical Physics, University of Oxford, Oxford OX1 3PU, United Kingdom}
\ead{$^{*}$benoit.mahault@ds.mpg.de}
\ead{$^{\dagger}$ramin.golestanian@ds.mpg.de}

\vspace{10pt}

\begin{indented}
\item[] \today
\end{indented}

\begin{abstract}
Point-like topological defects are singular configurations that occur in a variety of in and out of equilibrium systems with two-dimensional orientational order.
As they are associated with a nonzero circuitation condition, the presence of defects induces a long-range perturbation of the orientation landscape around them.
The effective dynamics of defects is thus generally described in terms of quasi-particles interacting through the orientation field they produce, whose evolution in the simplest setting is governed by the diffusion equation.
Due to the multivalued nature of the orientation field, its expression for a defect moving with an arbitrary trajectory cannot be obtained straightforwardly and is often evaluated in the quasi-static approximation.
Here, we instead propose an approach that 
allows to derive the exact expression for the orientation created by multiple moving defects, which we find to depend on their past trajectories and thus to be nonlocal in time.
Performing various expansions in relevant regimes, we show how improved approximations with respect to the quasi-static defect solution can be obtained.
Moreover, our results lead to so far unnoticed structures in the orientation field of moving defects which we discuss in light of existing experimental results.
\end{abstract}

\pagebreak

\tableofcontents

\title[The multivalued solution of the diffusion equation]{}

\begin{center}
   \rule{10cm}{1pt} 
\end{center}

\setcounter{page}{1}

\section{Introduction}

\label{sec_intro}

Topological defects are intimately associated with the emergence of spontaneous symmetry breaking, and the fundamental role they play in condensed matter systems is now well established.
Indeed, point-wise topological defects are commonly found in 
a wealth of two-dimensional systems such as 
passive~\cite{deGennes_Prost,Review_LC_Harth2020} and active~\cite{Shankar2022_toporeview,Thampi2013} liquid crystals, 
melting solids~\cite{StrandburgRMP1988}, turbulent fluids~\cite{Kraichnan1980turbulence},
thin superfluid films~\cite{AmbegaokarPRB1980,AgnoletPRB1989},  superconductors~\cite{ResnickPRL1981,HadzibabicNature2006} or trapped quantum gases~\cite{HadzibabicNature2006}.
Defects are known to be the main driver of the coarsening dynamics following a quench in a variety of passive~\cite{Bray2002review} and active~\cite{RanaPRE2020,Uchida2010,ChardacPRX2021,SahaPRX2020} systems.
At equilibrium, defect unbinding is also responsible for the celebrated Berezinskii-Kosterlitz-Thouless topological phase transition~\cite{berezinskii1971destruction,Kosterlitz1973JPC} that has been characterized in many of the systems cited above.
Because they continuously dissipate energy at the microscopic scale, active nematics~\cite{DoostmohammadiActNem} moreover self-organize into a chaotic `turbulent' phase~\cite{Alert2022turbulence,Berta2021} whose dynamics is driven by the steady creation and annihilation of defects.
Lastly, recent studies moreover suggest that the role of defects extends to biology as they may play a regulatory role in living tissues~\cite{Kawaguchi2017nature,Saw2017Nature}.

In many of the situations outlined above, it has been argued that defects can be described as quasi-particles interacting through the surrounding order parameter field~\cite{PismenBook}.
As the presence of defects introduces a long-ranged perturbation of the order with respect to the stationary uniform configuration, the resulting landscape acts as an effective force that sets the defects into motion.
Determining the shape of the order parameter field created by a defect is therefore central to understand the global behavior of the system
and has, even until recently, attracted much attention 
~\cite{Dubois-Violette1983,RodriguezPRA1991,DennistonPRB1996,Gartland2002ws,Najafi2003EPJB,Radzihovsky2015PRL,tang2017orientation,Cortese2018PRE,ShankarPRL2018,TangSM2019,Vafadefects2020,ZhangPRE2020,AnghelutaNJP2021}. 
In most practical cases, the norm of the order parameter relaxes fast enough such that only its local orientation is relevant to the dynamics of defects.
A broad range of theories moreover predict that the order orientation evolves according to the diffusion equation~\cite{PismenBook}. 
Even in the presence of additional nonlinear terms, e.g.\ due to coupling with an external flow or activity~\cite{TothPRL2002,Thampi2014EPL}, solutions of the diffusion equation still play an important role as they can serve as a basis for perturbation theory. 

Despite its apparent importance, to our knowledge, no explicit solution of the diffusion equation has been derived for a defect moving along an arbitrary trajectory.
Indeed, most studies focusing on characterizing the dynamics of defects rely on a quasi-static approximation where the orientation field is calculated 
from Laplace equation assuming immobile defects~\cite{Review_LC_Harth2020},
leading to the well-known Coulomb-type interaction forces between defects.
Some works have gone beyond the quasi-static approximation,
mostly considering defects moving with constant velocity~\cite{DennistonPRB1996,Radzihovsky2015PRL}.
This approximation, however, leads to unphysical divergences of the solution at large scales.
For rectilinearly moving defects, Rodriguez et al.~\cite{RodriguezPRA1991} further showed that defect acceleration leads to noticeable corrections in the resulting orientation field.

In this work, we propose a derivation of the orientation field created by point defects moving along arbitrary trajectories, in a similar vein to what has been previously done for defects interacting with elastic waves \cite{Najafi2003EPJB}. Contrary to most of the commonly employed approximations, the exact solution we derive is nonlocal in time and thus highlights the importance of the memory of past defect configurations to describe the dynamics of the system. 
In particular, we show that taking defect dynamics into account leads to new structures in the orientation field solution that are absent in the immobile defect limit. 
Our predictions are directly relevant to the dynamics of passive and active liquid crystals.

The remaining of this paper is organized as follows:
Section~\ref{sec_main_results} provides a brief introduction to the problem as well as the main steps leading to the dynamical defect solution of the diffusion equation.
Additionally, Sec.~\ref{sec_expansions} describes expansions of the single-defect solution for relevant regimes while
Sec.~\ref{sec_multi_defects} discusses the implications of our results to systems presenting multiple defects. 
Finally, we summarize our results and provide 
concluding remarks in Sec.~\ref{sec_discussion}.

\section{The orientation field generated by a moving defect}

\label{sec_main_results}

\subsection{Topological defects in the two-dimensional
Ginzburg-Landau equation}

\label{sec_GL_defects}

The most natural framework to describe $SO(2)$ spontaneously symmetry broken phases is the Ginzburg-Landau equation, 
which for an order parameter $\bphi(\x,t)$ follows
\begin{equation}
\label{RelGL}
    \partial_t\bphi(\x,t) = D\nabla^2\bphi(\x,t) + \chi(1-|\bphi(\x,t)|^2)\bphi(\x,t) ,
\end{equation}
where the parameters $D$ and $\chi$ are phenomenological. 
\colrev{For instance, a minimal description of polar and nematic liquid crystals is achieved by Eq.~\eref{RelGL} where $\bphi$ represents respectively the polar and nematic order parameters:
\begin{equation*}
    \bphi_{\rm polar} = \rho
    \left(\begin{array}{c} 
    \cos(\theta) \\ \sin(\theta)
    \end{array}\right), \qquad
    \bphi_{\rm nematic} = \rho
    \left(\begin{array}{c c} 
    \cos(2\theta) & \sin(2\theta) \\ 
    \sin(2\theta) & \cos(2\theta)
    \end{array}\right),
\end{equation*}
where $\rho(\x,t)$ and $\theta(\x,t)$ respectively set the magnitude and orientation of order, while $|\bphi|^2$ is a shorthand notation for the sum of the squared components of $\bphi$. 
$D$ in Eq.~\eref{RelGL} plays the role of an effective diffusivity and $\chi^{-1}$ sets the typical relaxation timescale of $\rho(\x,t)$.}
In particular, for $\chi \to \infty$ the norm $\rho$ is instantaneously forced to its equilibrium value $\rho = 1$. The dynamics of $\bphi$ then reduces to that of its orientation, which takes the simple form 
\begin{equation}
    \label{Diffusioneq}
    \partial_t\te(\x,t)=D\nabla^2\te(\x,t) .
\end{equation}

This picture is however incomplete, as in two dimensions and for generic initial and boundary conditions the field $\bphi$ may not uniformly converge to the ordered state $\rho(\x,t) = 1$, but present topological defects.
These configurations correspond to singularities of the orientation field $\te(\x,t)$ at a set of space points $\q_i(t)$ ($i = 1,\ldots$) and are by continuity of the order parameter admissible solutions of Eq.~\eref{RelGL} only under the condition $\rho(\q_i(t),t) = 0$ for all $i$.
All defect solutions carry a charge $s$ such that $2\pi s$ equals the circuitation of $\nabla\te(\x,t)$ along any closed curve encircling the defect center. 
Because of total charge conservation, defects are topologically constrained such that they must be created and annihilate in pairs with opposite charge.
For large $\chi$, the condition $\rho(\x,t)=1$ nevertheless remains true almost everywhere\footnote{Formally, $(D/\chi)^{1/2}$ defines a typical scale over which $\rho(\x,t)$ substantially differs from one around the defects. Outside of this `core' region, the dynamics of $\bphi$ is well captured by that of its orientation.}, such that the dynamics of a system with topological defects is fully characterized by that of the defects positions $\q_i(t)$ and of the orientation field $\te(\x,t)$.

\subsection{The multivalued solution of the diffusion equation}

\label{sec-sol-diffeq}

In this section, we derive the central result of this work, namely the general solution of the diffusion equation~\eref{Diffusioneq} for a defect moving along an arbitrary trajectory.
Otherwise stated, to lighten notations we work in what follows with time units such that the diffusivity $D$ is set to one without loss of generality. 

As the orientation of $\bphi$ is defined up to a constant phase shift --e.g.\ a multiple of $2\pi$ or $\pi$ for respectively ferromagnetic and nematic orders-- $\te(\x,t)$ is a multivalued function.
In particular, because of the nonzero circuitation condition a $s$-charged defect solution at position $\q$ imposes a $\pm 2\pi s$ discontinuous jump of $\te(\x,t)$ across an arbitrary branch cut extending from $\q$ to infinity.
For example, the orientation field $\thst(\x-\q)$ generated by a static defect satisfies $\nabla^2 \thst = 0$ and can be expressed as $\thst(\x-\q) = s\, {\rm arg}(\x - \q)$, which corresponds to a choice of cut such that $\thst \in (-\pi;\pi]$.
From this solution, it is then straightforward to calculate the physical gradient which must remain independent of the choice of the cut, namely
\begin{equation} \label{eq_static_nabla}
    \nabla\thst(\x) = s \eps \frac{\x}{|\x|^2}, \qquad 
    \eps \equiv \left( \begin{array}{cc} 
    0 &  -1\\
    1  & 0
    \end{array} \right) ,
\end{equation}
where $\eps$ is the two-dimensional antisymmetric Levi-Civita tensor which here corresponds to the $\frac{\pi}{2}$-rotation matrix.

For the general case of moving defects, however, one does not have an explicit expression for $\theta(\x,t)$ such that regularizing the space and time derivative operators is less straightforward~\cite{Kleinertbook}.
To circumvent the difficulties arising from the discontinuity of $\theta$ at the cut, we thus use the linearity of Eq.~\eref{Diffusioneq} and write the solution for a moving defect as
$
    \te(\x,t) = \thst(\x - \q(t)) + \varphi(\x,t),
$
where $\q(t)$ may now be time-dependent, while the remaining contribution $\varphi(\x,t)$ has zero curl and is thus smooth for all $\x \ne \q(t)$.
Defining $\vr \equiv \x - \q(t)$, we express Eq.~\eref{Diffusioneq} in the defect frame and find that $\varphi(\vr,t)$ solves
\begin{equation} \label{eq_diffeq_phi}
    \left[ \partial_t - \vi(t) \cdot \nabla - \nabla^2 \right] \varphi(\vr,t) = \vi(t) \cdot \nabla\thst(\vr),
\end{equation}
where $\vi(t) \equiv \dot{\q}(t)$ denotes the instantaneous velocity of the defect.
To solve Eq.~\eref{eq_diffeq_phi}, we consider the Green's function $G(\vr,t,t')$ whose evolution is governed by
\begin{equation}
    \label{refFT}
    \left[\partial_t - \vi(t)\cdt\nabla - \nabla^2 \right] G(\vr,t,t')=\delta^2(\vr)\delta(t-t').
\end{equation}
\colrev{Assuming uniform orientation at infinity,} solving Eq.~\eref{refFT} is easily achieved in Fourier space, leading to
\begin{equation}
\label{greenfunct}
    G(\vr,t,t')=\frac{\Theta(t-t')}{4\pi(t-t')}\exp\left[-\frac{|\vr+\Dq|^2}{4(t-t')}\right],
\end{equation}
where $\Theta(t-t')$ is the Heaviside step function and $\Dq\equiv\q(t)-\q(t')$.
Note that, as Eq.~\eref{eq_diffeq_phi} is in general not invariant under time translations, the Green's function~\eref{greenfunct} depends separately on $t$ and $t'$, and not just on the difference $t-t'$.
The general solution of Eq.~\eref{eq_diffeq_phi} is therefore given by
\begin{equation} \label{eq_phi_explicit}
    \varphi(\vr,t) = \int \rmd^2y\int\rmd t' \, G(\vr - \y,t,t') (\vi(t')\cdot \nabla)\thst(\y).
\end{equation}

For most applications in the study of topological defects, the gradient of the angular field $\te$ is actually a more useful quantity than $\te(\vr,t)$ itself. 
In what follows, we thus calculate $\nabla\varphi$ and show that, contrary to $\varphi$, its expression can be simplified into a form that is local in space.
As the Green's function $G(\vr-\y,t,t')$ satisfies $\nabla_{\vr}G = -\nabla_{\y}G$, we obtain from~\eref{eq_phi_explicit} after integration by parts
\begin{eqnarray}
    \nabla\varphi(\vr,t) & = & -\eps\int_{{\cal C}_a}\rmd\bi{l}\int\rmd t' \, G(\vr - \y,t,t') (\vi(t')\cdot \nabla)\thst(\y) \nonumber \\
    & & + \int\rmd^2y\int\rmd t' \, G(\vr - \y,t,t') ( \vi(t')\cdot \nabla)\nabla\thst(\y),
    \label{eq_nabla_phi1}
\end{eqnarray}
where the boundary contribution on the first line was retained due to the singular behavior of $\nabla\thst(\y)$ at $\y = \bf{0}$.
To evaluate it, we thus consider a circle ${\cal C}_a$ of radius $a \to 0$ around the defect.
After some algebra detailed in~\ref{app_nablathe}, we find that this term indeed leads to a nonvanishing contribution for $a = 0$ given by $\pi s \eps \int\rmd t' \vi(t') G(\vr,t,t')$.
Integrating by parts the remaining term in~\eref{eq_nabla_phi1} then leads to a similar boundary contribution, such that we obtain
\begin{eqnarray}
    \nabla\varphi(\vr,t) & = & -2\pi s \eps  \int\rmd t' \, \vi(t') G(\vr,t,t') \nonumber \\
    & & + \int\rmd^2y\int\rmd t' \, ( \vi(t')\cdot \nabla_{\vr})G(\vr - \y,t,t') \nabla\thst(\y),
    \label{eq_nabla_phi2}
\end{eqnarray}
were on the second line we indicate explicitly that the gradient of $G$ is taken with respect to the variable $\vr$ in order to avoid possible confusion.
To evaluate the spatial integral on the second line, we note that the Green's function satisfies the identity
$(\partial_t - \vi(t)\cdot\nabla)G(\vr,t,t')
= -(\partial_{t'} + \vi(t')\cdot\nabla)G(\vr,t,t')$,
which can be checked from~\eref{greenfunct} by direct calculation.
Using moreover Eq.~\eref{refFT}, we get
\begin{equation}\label{ident}
    (\vi(t')\cdot\nabla)G(\vr,t,t') = (\partial_{t'} + \nabla^2)\left[G_{\rm D}(\vr,t-t') - G(\vr,t,t')\right],
\end{equation}
where 
\begin{equation*}
    G_{\rm D}(\vr,t-t') \equiv \frac{\Theta(t - t')}{4\pi(t-t')}e^{-\frac{|\vr|^2}{4(t-t')}},
\end{equation*}
is the Green's function of the diffusion equation.
Replacing~\eref{ident} into Eq.~\eref{eq_nabla_phi2}, the $\propto \partial_{t'}$ terms cancel upon integration, while the $\propto \nabla^2$ terms are calculated via additional integration by parts keeping the associated boundary contributions from the defect core.
After some calculations detailed in~\ref{app_nablathe}, we obtain
\begin{equation} \label{eq_nabla_phi_final}
    \nabla\varphi(\vr,t) = -2\pi s \eps  \int\rmd t' \, \left[\left(\nabla + \vi(t')\right) G(\vr,t,t') - \nabla G_{\rm D}(\vr,t-t')\right].
\end{equation}
Noting that the static defect solution satisfies $\nabla\thst(\vr) = -2\pi s \eps \int\rmd t' \nabla G_{\rm D}(\vr,t-t')$,
we finally get after explicitly replacing $G$ by its expression
\begin{equation} \label{refgrad}
    \fl \qquad \nabla\theta(\vr,t) = \nabla[\thst(\vr) + \varphi(\vr,t)] = -\frac{s}{2} \eps  \int_{-\infty}^t \frac{\rmd t'}{(t-t')} \, \left(\nabla + \vi(t')\right) e^{-\frac{|\vr -\Dq|^2}{4(t-t')}}.
\end{equation}
Strikingly, the r.h.s.\ of~\eref{refgrad} cannot be written as the gradient of a scalar field.
This feature is a consequence of the fact that $\theta$ is a multivalued function.
In fact, we show in~\ref{app_nablathe} that $\nabla\theta(\vr,t)$ is irrotational everywhere except at the defect position: $\left[\nabla \times \nabla\theta(\vr,t)\right]\cdot\hat{\z} = 2\pi s \delta^2(\vr)$.
Consequently, for any closed curve $\Upsilon$ oriented along the counter-clockwise direction, it follows from Green's theorem that
\begin{equation}
\label{circcond}
    \oint_{\Upsilon}\rmd\bi{l}\cdot \nabla\theta({\bi l},t) = 2\pi s\int_{S(\Upsilon)}\rmd^2r\,\delta^2(\vr)=
    \cases{
        2\pi s & ${\rm if}\; {\bf 0} \in S(\Upsilon)$ \\
        0 & {\rm otherwise}
    } ,
\end{equation}
where $S(\Upsilon)$ denotes the surface enclosed by $\Upsilon$.

The r.h.s.\ of Eq.~\eref{refgrad} is nonlocal in time, which highlights that the orientation field generated by a moving defect depends on the full history of its trajectory.
Nevertheless, we will show in Sec.~\ref{sec_expansions} that in a number of limiting regimes the expression~\eref{refgrad} can be approximated by forms which are local in time.
In these cases, the time dependency of the orientation is indeed due only to the instantaneous defect position and velocity.
A straightforward example is obtained fixing $\q(t) = \q$ and $\vi(t) = \bf 0$ in  Eq.~\eref{refgrad}, 
leading as expected to $\nabla\theta(\vr,t) = \nabla\thst(\vr)$.
For motile defects, on the other hand, the past history of the trajectory still generally appears effectively through the value of some coefficients.
A particularly enlightening case is addressed in Sec.~\ref{sec_far_field} where the far field limit of Eq.~\eref{refgrad} is derived.
There, we show that in addition to the static defect solution~\eref{eq_static_nabla} an additional contribution given by the average defect angular momentum arises.
This contribution is moreover orthogonal to~\eref{eq_static_nabla}, such that the solution~\eref{refgrad} predicts that a defect with nonzero angular momentum generates spiraling field lines,
as experimentally observed in a number of liquid crystal systems~\cite{RotatingDefs,MissaouiPRR2020}.

In some cases, it will be more convenient to work with $\nabla\theta$ expressed in the lab frame, which is recovered using $\x = \vr + \q(t)$ in~\eref{refgrad}:
\begin{equation}
    \label{labgrad}
    \nabla\theta(\x,t)=-\frac{s}{2}\eps \int_{-\infty}^{t} \frac{\rmd t' }{(t-t')} \left[\nabla+\vi(t')\right]e^{-\frac{|\x-\q(t')|^2}{4(t-t')}}.
\end{equation}
In particular, we will show in Sec.~\ref{sec_multi_defects} how the angular field landscape generated by multiple defects is trivially obtained from~\eref{labgrad} by summation of the single defects solutions.
Since the exact solution accounts for the full history of the system, it allows us to derive in Sec.~\ref{sec_multi-defects-annihilation} the functional form of the propagation of the orientation perturbation following the creation of a defect pair, as well as that of the field relaxation after the annihilation of the pair.
Our results moreover point out that the relaxation of $|\nabla\theta|$ to zero will take different scaling forms depending on the ratio of the time delay since annihilation and the total life time of the defect pair. 
Finally, we show in Sec.~\ref{sec_multi-defects-mismatch} that the angular momentum dependency of the far field limit of Eq.~\eref{labgrad} generically leads to defect pair solutions with an angular field that cannot be described as the sum of two static defect configurations, and are known as mismatched configurations~\cite{tang2017orientation,GiomiOrientation,MissaouiPRR2020}.
This result suggests that in complex environments with many interacting defects mismatching shall naturally occur even without explicit elastic anisotropy or specifically imposed boundary conditions.
 
\section{Some useful expansions}

\label{sec_expansions}

The results derived in the previous section are exact and valid on all the plane and for all times, but due to its non-locality in time the solution~\eref{refgrad} is in practice of limited use.
In this section, we therefore investigate relevant limiting cases for which Eq.~(\ref{refgrad}) simplifies into more manageable expressions.

\subsection{The near field expansion}

\label{sec_near_field}

Here, we show by performing a near field expansion of Eq.~\eref{refgrad} that the expression of the orientation gradient simplifies in the vicinity of the topological defect.
Knowing the structure of the field around a defect is useful in practice to calculate the dynamics of its position $\q(t)$, as it is in particular involved in the expression of the defect mobility~\cite{PismenBook,DennistonPRB1996}.
Moreover, as $\nabla\theta(\x,t)$ essentially sets the interaction between defects, our result highlights its dominating contribution for nearby defects.

As the diffusion equation~\eref{Diffusioneq} does not have any intrinsic scale, we cannot write the desired expansion in terms of a small length parameter. However, the defect position $\vr = \bf 0$ corresponds by construction to a singularity of the orientation field gradient. 
In close proximity of the defect core, this diverging contribution thus naturally dominates the expression of $\nabla\theta(\vr,t)$.
More generally, in this subsection we will derive the contribution to the gradient~\eref{refgrad} that is discontinuous at the topological singularity. 
This contribution contains, but is not limited to, the dominating singular part.

Firstly, we  split the time integral in~\eref{refgrad} into two contributions respectively over $(-\infty;t-\tau)$ and $(t-\tau;t)$ with $\tau > 0$. The first contribution is clearly analytical in $\vr = \bf 0$, while the second is not.
In particular, it is easily shown that the non-analyticity of the second contribution at $\vr = \bf 0$ arises because of the $(t-t')^{-1}$ factor in the integral. 
Therefore, as we show now the near field singular part of the solution only depends on the defect configuration at time $t$.
Taking $\tau$ small, we Taylor expand around $t$ the defect velocity and displacement as \colrev{
\begin{equation*}
\vi(t')=\vi(t) + {O}(t'-t), \qquad
\Dq=\vi(t)(t-t') + {O}((t'-t)^2).
\end{equation*}
Substituting these expressions in~\eref{refgrad}, 
it is straightforward to show that the ${O}(t'-t)$ and ${O}((t'-t)^2)$ terms respectively for $\vi(t')$ and $\Dq$ do not lead to any singular contribution.}
We therefore formally write
\begin{equation*} \label{nablapart_nf_int}
    \nabla\theta(\vr,t) \underset{r\to 0}{=}
    -\frac{s}{2}\eps\int_{t-\tau}^{t} \frac{\rmd t'}{(t-t')} \left[\nabla+\vi(t)\right]e^{-\frac{r^2}{4(t-t')}-\frac{\vi(t)\cdot\vr}{2}} + {\rm c.t.} ,
\end{equation*}
where $r \equiv |\vr|$ and  ``${\rm c.t.}$'' refers to the subdominant continuous terms. 
The integral \colrev{in this expression} is in turn evaluated using the change of variables $u=r^2/(t-t')$.
Taking the limit $r \to 0$, we finally end up with
\begin{equation}
        \label{nearexp}
         \nabla\theta(\vr,t) \underset{r\to 0}{=} s\eps\left(\frac{\hat \vr}{r}+\frac{\vi(t)}{2}\ln\lt(\frac{r}{\lambda}\rt)-\frac{\vi(t)\cdot \hat \vr}{2}\hat \vr\right) + {\rm c.t.} ,
    \end{equation}
where $\hat \vr \equiv \vr / r$ and $\lambda$ is an unknown length scale introduced for dimensional reasons that we discuss further below.

The dominating term in the expansion~\eref{nearexp} is given by the static defect contribution $s\eps\hat \vr/r$.
Moreover, the motion of the defect leads to additional subdominant discontinuous terms which depend only on the instantaneous velocity $\vi(t)$.    
The dominating near field behavior of $\nabla\theta(\vr,t)$ is therefore always local in time, although the full solution~\eref{refgrad} is not.
Note that to keep the argument of the logarithm in~\eref{nearexp} nondimensional we have included a length scale parameter $\lambda$.
However, it is clear that $\lambda$ can in general be absorbed into the continuous contribution to the gradient.
Therefore and contrary to the other discontinuous terms, the scale $\lambda$ generally depends on the past positions and velocities of the defect, and cannot be computed by such a near field expansion. 
 
\subsection{The far field expansion}

\label{sec_far_field}

Having characterized the near field behavior of the solution~\eref{refgrad}, we now turn to the opposite limit of far field.
This limit is often used to model the interaction between defects, which for static defects is of Coulomb-like form~\cite{Review_LC_Harth2020}.
As we detail below, taking into account the dynamics of the defect we uncover a new contribution to the far field which is generated by the defect angular momentum.
This contribution, to our knowledge unreported so far, qualitatively modifies the orientation field generated by the defect.

As noted previously, the diffusion equation~\eref{Diffusioneq} does not carry any intrinsic length scale, such that to perform the expansion we must introduce one.
In this section we thus suppose that the defect moves inside a bounded region of space.
Hence, there exists a length scale $\ell$ such that the relative displacement $|\Dq|<\ell$ for all past times $t'$. 
As detailed in~\ref{app_far}, performing an expansion of the solution~\eref{refgrad} up to first order in $\ell/r$ we obtain 
\begin{equation}
\label{farfirst}
    \nabla\te(\vr,t)=s\eps\frac{\hat \vr}{r}
    +s\vr\int_{-\infty}^{t}\rmd t'\, \frac{L(t')}{8(t-t')^2}e^{-\frac{r^2}{4(t-t')}}+{\cal O}\left(\frac{\ell}{r}\right),
\end{equation}
where $L(t')\equiv v_1(t')\Delta q_2(t,t')-v_2(t')\Delta q_1(t,t')$ denotes the angular momentum of the defect at time $t'$.
As for the near field, the first term on the r.h.s.\ of Eq.~\eref{farfirst} corresponds to the solution expected for a static defect.  
The second term, on the contrary, bears a dynamical origin as it emerges when the defect spins. 
Applying the change of variable $u=r^2/(t-t')$ in the integral, it is moreover straightforward to show that the latter effectively scales as $r^{-2}$, such that for $L$ finite both contributions to the far field may have comparable amplitudes.
We also note that the new dynamical term is always radial, which ensures that the circuitation condition~\eref{circcond} is always satisfied. 
Finally, although the angular momentum in~\eref{farfirst} is computed with respect to the defect position at time $t$, any other choice $\q_c$ with $|\q(t) - \q_c| < \ell$ would lead to subdominant contributions $\sim {\cal O}(\ell/r)$, such that its dependency in $t$ is left implicit.

The expression~\eref{farfirst} can be simplified further assuming that the defect angular momentum averages to a finite value $\eL$ in the long-time limit:
\begin{equation}
    \eL \equiv \lim_{T\rightarrow\infty}\frac{1}{T+t}\int_{-T}^{t}\rmd t' \, L(t') .
\end{equation}
This scenario is for example relevant to the case where $L(t')$ oscillates around $\eL$ with a characteristic timescale $\tau_L$. 
Writing $L(t')=\eL+L_{\rm osc}(t')$ where $L_{\rm osc}(t')$ accounts for the oscillating part of the angular momentum, we rewrite the integral in~\eref{farfirst} as
\begin{eqnarray}
 \fl \int_{-\infty}^{t}\rmd t'\,\frac{L(t')}{8(t-t')^2}e^{-\frac{r^2}{4(t-t')}} & =\int_{-\infty}^{t}\rmd t'\,\frac{\eL}{8(t-t')^2}e^{-\frac{r^2}{4(t-t')}}+\int_{-\infty}^{t}\rmd t'\,\frac{L_{\rm osc}(t')}{8(t-t')^2}e^{-\frac{r^2}{4(t-t')}}\nonumber\\
 \label{avgmom} \fl  
 &=\frac{\eL}{2r^2} - \int_{-\infty}^{t}\rmd t' \, P(t')\partial_{t'}\left[\frac{1}{8(t-t')^2}e^{-\frac{r^2}{4(t-t')}}\right] ,
\end{eqnarray}
where the second equality was obtained integrating by parts and $P(t')$ is a primitive of $L_{\rm osc}(t')$.
Noting that $P(t') = {\cal O}(\tau_L)$ by construction, we conclude that the second term on the r.h.s.\ of Eq.~\eref{avgmom} is of order $\tau_L/r^2$.
Therefore, considering length scales $r \gg \tau_L^{1/2}$, Eq.~\eref{farfirst} simplifies at leading order to
\begin{equation}
    \label{fargrad}
    \nabla\te(\vr,t) \simeq s \eps \frac{\hat \vr}{r} + \frac{s \eL}{2} \frac{\hat \vr}{r} .
\end{equation}
Again, the orientation gradient caries a radial contribution $\propto \eL$ due to the defect motion whose amplitude decays as the inverse of the distance $r$ to the defect center, similarly to the tangential static contribution.

\begin{figure}[!t]
    \centering
    \includegraphics[width=.5\linewidth]{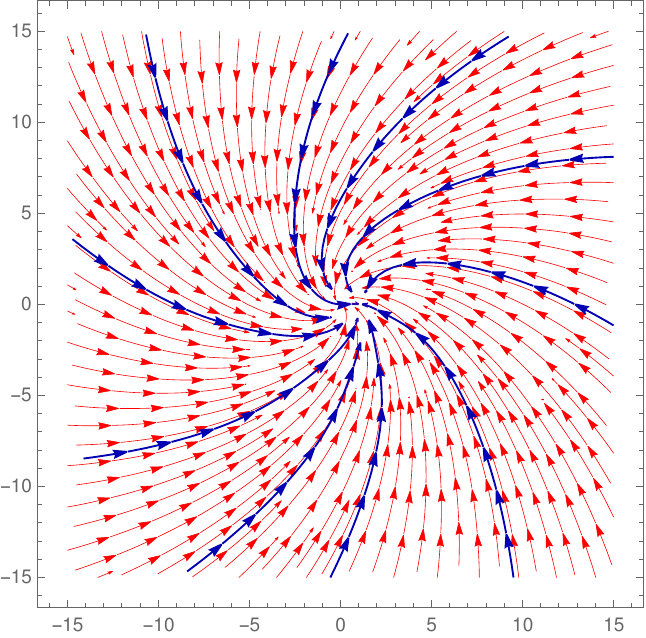}
    \caption{Streamlines of the effective force field $\eps\nabla\te$ for a defect moving at constant speed $v=1$ along a circular trajectory with radius $R=1$. The blue lines show the exact solution obtained by numerically integrating \eref{refgrad} while the far field approximation~\eref{fargrad} is shown in red.}
    \label{fig_spiral}
\end{figure}

As the effective force putting defects into motion is orthogonal to the gradient of the orientation field ($\bi{F}_{\rm eff}\propto\eps\nabla\te$)~\cite{DennistonPRB1996,PismenBook},
static defects essentially interact via Coulomb-type interactions leading to an elegant analogy with charged particles dynamics. 
The dynamical contribution to Eq.~\eref{fargrad}, on the contrary, leads to tangential (or solenoidal) forces between the defects.
This force is proportional to $m \equiv s \eL$,
which corresponds to the magnetic moment of a particle of charge $s$ moving along circular trajectories with associated angular momentum $\eL$. 
The analogy with charged particles is however limited, as the resulting angular momentum induced interaction differs from that of actual magnetic dipoles. 
Figure~\ref{fig_spiral} thus shows that the contribution of the defect angular momentum leads to a spiraling of the force field lines. As we will discuss further in Sec.~\ref{sec_multi-defects-mismatch}, oppositely-charged rotating defects will thus generally annihilate following curved trajectories.

Integrating \eref{fargrad} leads to the following expression for the angular field:
\begin{equation}
    \label{farang}
    \theta(\vr,t)=s\arg(\vr)+\frac{s \eL}{2}\ln\lt(\frac{r}{\lambda}\rt)+\te_0 ,
\end{equation}
where $\te_0$ is an integration constant and the length scale $\lambda$ was introduced for dimensional reasons. 
Despite the fact that Eq.~\eref{farang} describes the orientation far field generated by a moving charge, 
it takes a quasi-static form when expressed in the reference frame of the defect,
as it solves the Laplace equation: $\nabla^2\te(\vr,t)=0$.

We conclude this section by noting that spinning topological defects have recently been realized in sandwiched liquid crystal suspensions~\cite{RotatingDefs}.
Adding micro-rods to the suspension, the former are indeed attracted to the core of topological defects while the application of an alternating electric field along the third dimension results in a spinning motion of the rods, which drive the defects along circular trajectories.
Schlieren textures of rotating defects then clearly show a characteristic spiral shape with an orientation set by the chirality of the defect trajectory (clockwise or counter-clockwise).
To evaluate the strength of the angular momentum in~\eref{farang}, we note that it is expressed in units of the orientation diffusion coefficient $D$. 
Reference~\cite{RotatingDefs} reports defect circular trajectories with typical radius $R \simeq 7 \,\mu{\rm m}$ and speed  $v \simeq 7 \,\mu {\rm m}\,{\rm s}^{-1}$, leading to an angular momentum of the order of $\eL \simeq 50 \,\mu {\rm m}^2\, {\rm s}^{-1}$.
As the value of $D$ is expected to lie in the range $10-150 \,\mu {\rm m}^2\,{\rm s}^{-1}$~\cite{RotationalViscosity}, we conclude that in this experiment $\eL/D$ can easily take values of order one, leading to a clear spiraling of director field lines around the defect.

\subsection{The low mobility expansion}

\label{sec_low_mob}

\subsubsection{The general expression for a slow defect}

\label{sec_nabla_slow_def}

\begin{figure}[!t]
\centering
\includegraphics[width=0.8\textwidth]{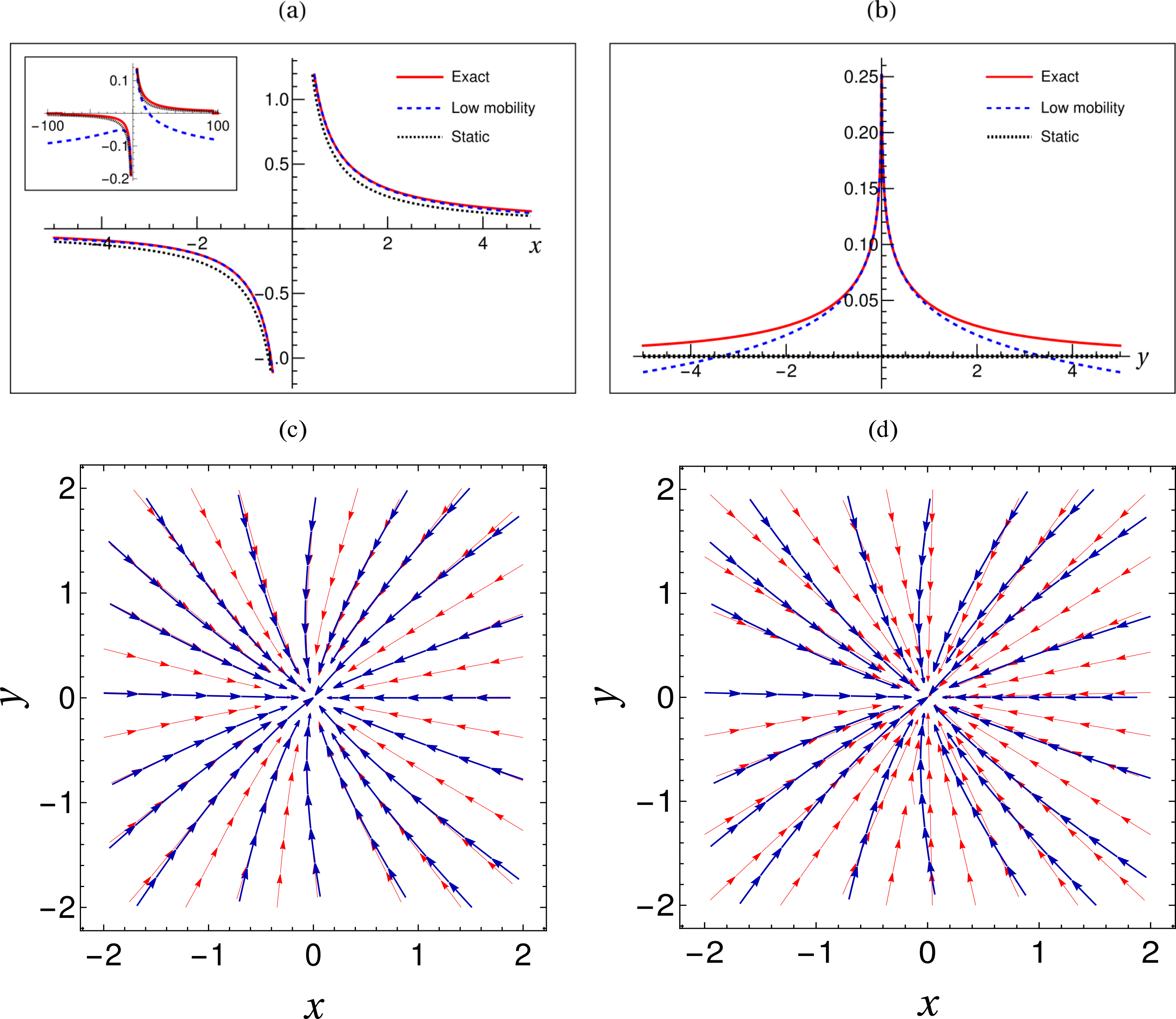}
\caption{Orientation field created by a defect moving along the $x$ axis following the trajectory defined by Eq.~\eref{coudyn}.
Panels (a) and (b) show the $y$ component of the gradient function in the defect comoving frame respectively as function of $x$ ($y = 0$) and $y$ ($x = 0$). 
The inset if (a) shows the same data over a broader $x$ range.
The red continuous lines show the exact solution computed with~\eref{refgrad}, while the dashed blue and black curves correspond respectively to the low mobility expansion~\eref{coulowmob} and static approximation~\eref{eq_static_nabla}.
\colrev{Panels (c) and (d) respectively compare streamlines of the effective force field $\eps\nabla\te$ obtained in the comoving frame from the exact solution (in blue) and the low mobility(c) and static(d)  approximation (in red). 
In all the panels the mobility is set to $\mu = 1/\ln(100)$ and the fields are evaluated at $q = 1$.}
}
\label{axisplot}
\end{figure}

For the third and last expansion, we focus on slowly moving defects. The position of a defect indeed generally evolves in an overdamped manner as~\cite{PismenBook}
\begin{equation}
\label{damped}
    \dot{\q}(t) = \mu \bi{F}_{\rm eff},
\end{equation}
where $\mu$ is an effective mobility and $\bi{F}_{\rm eff}$ denotes the sum of effective forces applied to the defect, which in our system of units has the dimension of an inverse length. $\bi{F}_{\rm eff}$ can either result from the application of an external drive such as an electric field, or from the orientation field of other defects.
Estimates of the defect mobility show that it scales as $\mu^{-1} \sim \ln(\lambda/a)$ where $\lambda$ is a macroscopic scale of the problem (e.g. the system size) and $a \to 0$ typically sets the size of the defect core~\cite{Frictcoeff,PismenBook}.  
$\mu$, here expressed in units of the diffusivity $D$, is thus generally considered small such that most studies adopt the quasi-static approximation for which the orientation field of the defect is evaluated from the static solution ($\dot{\q}(t) = \bf 0$). 
This approximation, which corresponds to a zeroth order expansion of~\eref{refgrad} in terms of $\mu$, is useful in practice but might break down in cases where the velocity of the defect is not negligible, as shown in Fig.~\ref{axisplot}.
Here, we therefore evaluate the first order correction terms to the static solution by expanding~\eref{refgrad} up to ${\cal O}(\mu)$.

Because of the presence of the $\mu$ factor on the r.h.s.\ of Eq.~\eref{damped}, the $n^{\rm th}$ time derivative $\q^{(n)}(t)$ of $\q(t)$ is of order $\mu^n$.
Therefore, we can write up to order $\mu^2$
\begin{equation}
\label{trajexp}
\q(t')=\q(t)+\vi(t)(t'-t)+{\cal O}(\mu^2), \qquad \vi(t')=\vi(t)+{\cal O}(\mu^2).
\end{equation}
These expansions, however, do not hold in the long time limit where $t - t' \to \infty$.
Indeed, substituting~\eref{trajexp} into the solution~\eref{refgrad} result in a diverging integral for $t' \to -\infty$.
We then regularize the integral by introducing a parameter $\varepsilon$ and consider instead of~\eref{refgrad}
\begin{equation}
    \label{epsgrad}
    \bi{K}_\varepsilon(\vr,t) \equiv -\frac{s}{2}\eps \int_{-\infty}^{t} \rmd t'\, \left[\nabla+\vi(t')\right]\left[\frac{\xi^{\varepsilon}(t')}{(t-t')^{1+\varepsilon}}e^{-\frac{|\vr+\Dq|^2}{4(t-t')}}\right] ,
\end{equation}
where $\xi(t)$ is a timescale introduced for dimensional reasons, and it is clear that 
$\nabla\theta(\vr,t) = \lim_{\varepsilon\to0}\bi{K}_\varepsilon(\vr,t)$.
Crucially, for all $\varepsilon>0$ the integral~\eref{epsgrad} converges even under the approximation~\eref{trajexp}.
Namely, we find up to ${\cal O}(\mu^2)$ terms that
\begin{equation}
   \bi{K}_\varepsilon(\vr,t) = -\frac{s}{2}\eps\left[\nabla+\vi(t)\right]\left[\left(1-\frac{\vi(t)\cdt\vr}{2}\right)\left(\frac{4\xi(t)}{r^2}\right)^{\varepsilon}\Gamma(\varepsilon) \right] ,
\end{equation}
where $\Gamma$ denotes the standard Gamma function. 
Expanding this expression for small $\varepsilon$, 
we moreover get
\begin{equation}
\label{diverg}
 \bi{K}_\varepsilon(\vr,t) = s\eps\left( \frac{\hat\vr}{r}+\frac{\vi(t)}{2}\ln\left(\frac{e^{\gamma_{\rm E}/2}r}{2\sqrt{\xi(t)}}\right) -\frac{\vi(t)\cdt\hat \vr}{2} \hat \vr -\frac{\vi(t)}{4\varepsilon} \right) + {\cal O}(\varepsilon) ,
\end{equation}
where $\gamma_{\rm E}$ stands for the Euler-Mascheroni constant and we used the approximation of the Gamma function $\Gamma(\varepsilon)= \varepsilon^{-1} - \gamma_{\rm E} + {\cal O}(\varepsilon)$. 

Unsurprisingly, the expansion~\eref{diverg} includes a divergent term $\sim \varepsilon^{-1}$.
This term is unphysical and due to the fact that the low mobility expansion~\eref{trajexp} breaks down at long times.
To regularize $\bi{K}_\varepsilon(\vr,t)$, we note that the function $\alpha(\vr,t)\equiv\kappa \vr\cdt(\eps\vi(t))$ is a solution of the diffusion equation (up to ${\cal O}(\mu^2)$ terms) for arbitrary $\kappa$, as $\nabla^2\alpha(\vr,t)=0$ and $\partial_t\alpha(\vr,t)={\cal O}(\mu^2)$.
Hence, from the superposition principle the field $\theta(\vr,t)$ satisfying $\nabla\theta(\vr,t) = \lim_{\varepsilon \to 0} \bi{K}_\varepsilon(\vr,t) + \nabla\alpha(\vr,t)$ is also a solution of the diffusion equation~\eref{Diffusioneq}.
Taking $\kappa = s/(4\varepsilon)$, we can thus regularize Eq.~\eref{diverg} which for $\varepsilon = 0$ leads to
\begin{equation}
    \label{lowmob}
    \nabla\te(\vr,t) = s\eps \left( \frac{\hat \vr}{r}+\frac{\vi(t)}{2}\ln\left(\frac{e^{\gamma_{\rm E}/2}r}{2\sqrt{\xi(t)}}\right) -\frac{\vi(t)\cdt\hat \vr}{2}\hat \vr \right),
\end{equation}
up to terms of order $\mu^2$.

Comparing, Eqs.~\eref{nearexp} and~\eref{lowmob}, we find that the near field and low mobility expansions take similar expressions. 
However, contrary to Eq.~\eref{nearexp} the undetermined contribution to the expression~\eref{lowmob} is not a continuous function, but simply a uniform (in space) timescale $\xi(t)$.
The value of $\xi(t)$ depends on external factors such as the specific form of the equation of motion~\eref{damped},
but it can generally be obtained by computing, analytically or numerically, the exact value of the gradient~\eref{refgrad} at a given point in space $\vr^*$. 
Similarly, since the difference $\bi{K}_\varepsilon(\vr,t) - \bi{K}_\varepsilon(\vr^*,t)$ is convergent for $\varepsilon \to 0$ and its limit is independent of $\xi$,
the value of the approximation~\eref{lowmob} can be calculated at every point $\vr$ by adding $\lim_{\varepsilon\to0}\bi{K}_\varepsilon(\vr,t) - \bi{K}_\varepsilon(\vr^*,t)$ to the value of the integral~\eref{refgrad} at $\vr^*$.

To quantify the
disturbances of the orientation field created by the motion of the defect, 
we integrate Eq.~\eref{lowmob} and get
\begin{equation}
    \label{theta_lowmob}
    \te(\vr,t) = s\arg[\vr] + \frac{s }{2} 
    \left( \ln\left(\frac{e^{\gamma_{\rm E}/2}r}{2\sqrt{\xi(t)}}\right) - 1 \right) \vr\cdt[\eps \vi(t)] + \theta_0,
\end{equation}
where $\theta_0$ is a constant of integration.
Looking at the order orientation profile along a circle centered in $\vr = \bf 0$, Eq.~\eref{theta_lowmob} predicts that the motion of the defect leads at leading order to a sinusoidal perturbation with respect to the static contribution $s \arg[\vr]$. 
Interestingly, similar sinusoidal profiles have been reported in the chaotic phase of a two-dimensional active nematics where defects are strongly motile~\cite{He2019PNAS}.

\subsubsection{An explicit expression for defects interacting via Coulomb forces}

\label{sec_theta_coulomb}

To exemplify how the approximation~\eref{lowmob} can be applied in practice, we consider the case where the equation of motion of the defect follows
\begin{equation}
\label{coudyn}
\dot{\q} = -\mu \hat \q /q + {\cal O}(\mu^2).
\end{equation}
This choice is of course motivated, as Eq.~\eref{coudyn} is relevant to the annihilation dynamics of two oppositely charged defects. 
As noted before, the force acting on a defect is orthogonal to the gradient of the orientation field landscape due to the presence of other defects, 
while at zeroth order in the mobility the perturbation due to the presence of a defect is given by the static expression \eref{eq_static_nabla}.
Moreover, in this case the center of mass of the two defects remains immobile and can be set at $\bf 0$ without loss of generality, such that Eq.~\eref{coudyn} indeed corresponds to the expected equation of motion for a pair of defects at first order in $\mu$. 
Taking an initial condition for $\q$ aligned with the unit vector $\uveci$, the solution of Eq.~\eref{coudyn} which passes through the point $\q(t)$ at time $t$ reads
\begin{equation}
    \label{trajeq}
    \q(t')=\sqrt{q^2(t)+2\mu (t-t')} \, \uveci.
\end{equation}
As stated previously, to get a closed form of the low mobility expansion~\eref{lowmob} we have to evaluate the value of the gradient~\eref{refgrad} at a particular point $\vr^*$.
Here, a natural choice is the location $\x = \bf 0$ of the center of mass between the defects, which in the comoving coordinate frame corresponds to $\vr^* = -\q(t)$.
We can thus fix the value of $\xi(t)$ by numerically evaluating the integral in~\eref{refgrad} at $\vr^*$ for all $t$.
In the particular case where $\q(t')$ is given by the solution~\eref{trajeq}, we can instead proceed analytically. 
Let us consider Eq.~\eref{refgrad} with $\vr = -\q(t)$.
The resulting time integral can be calculated exactly, namely we find up to order $\mu^2$
\begin{equation}
\label{specvalue}
    \nabla\te(-\q(t),t)=-\frac{s}{q}\eps\uveci\left(1-\frac{\mu}{2}-\frac{\mu}{4}\ln\left(\frac{8e^{-1-\gamma_{\rm E}}}{\mu}\right)\right),
\end{equation}
while calculation details are presented in~\ref{app_lowmob-xi}.
Comparing Eq.~\eref{specvalue} with Eq.~\eref{lowmob} for $\vr = -\q(t)$ and $\vi(t) = -\mu \uveci / q(t)$, the two expressions coincide for $\xi(t) = 2q(t)/(e\mu)$ such that we finally obtain the closed form
\begin{equation}
    \label{coulowmob}
    \nabla\te(\vr,t)=s\eps
    \left( \frac{\hat \vr}{r}
    +\frac{\vi(t)}{2}\ln\left(\frac{\sqrt{\mu}e^{\frac{\gamma_{\rm E}+1}{2}}r}{2\sqrt{2}q(t)}\right) -\frac{\vi(t)\cdt\hat \vr}{2}\hat \vr \right).
\end{equation}

Evaluating numerically the exact solution~\eref{refgrad} for the trajectory~\eref{trajeq}, we now verify how well the low mobility approximation~\eref{coulowmob} performs.
The only free parameter to fix for the comparison is the value of the mobility $\mu$. 
Since $\mu$ typically decays as the inverse of the logarithm of the defect core radius~\cite{PismenBook}, it is generally not extremely small in realistic scenarios. For the present application, we choose in particular $\mu^{-1} = \ln(100) \simeq 4.6$.

\begin{figure}[t!]
\centering

    \includegraphics[width=\textwidth]{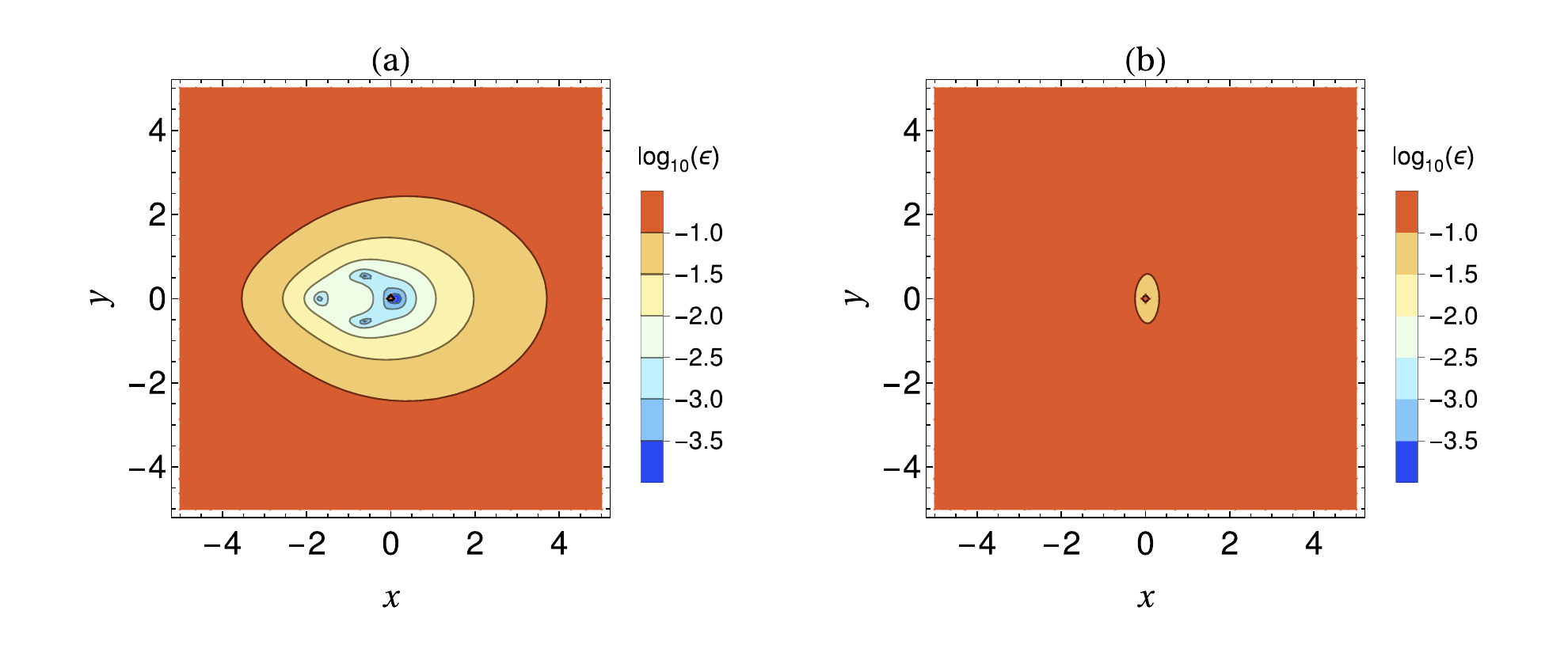}

\caption{Relative error comparing the exact gradient (Eq.~\eref{refgrad}) with the low mobility ((a), Eq.~\eref{lowmob}) and static ((b), Eq.~\eref{eq_static_nabla}) approximations for a
defect moving along the $x$ axis and following the trajectory defined by Eq.~\eref{coudyn}. 
In both panels the mobility is set to $\mu = 1/\ln(100)$ and the fields are evaluated at $q = 1$ while coordinates frame are chosen such that the defect is located at the origin.}
\label{error}
\end{figure}
As a first remark, we immediately see that because it diverges logarithmically for $r \to \infty$, the expansion~\eref{coulowmob} is not accurate far away from the defect regardless of the value of $\mu$.
On the contrary, Fig.~\ref{axisplot} shows that close enough from the defect center the first order low mobility expansion~\eref{coulowmob} convincingly approximates the exact solution~\eref{refgrad}.
These observations are confirmed by Fig.~\ref{error} which displays maps of the relative error
\begin{equation*}
    \varepsilon_{\rm rel}(\vr,t) \equiv \frac{|\nabla\te(\vr,t)-\nabla\te_{\rm app}(\vr,t)|}{|\nabla\te(\vr,t)|},
\end{equation*}
where $\nabla\te(\vr,t)$ and $\nabla\te_{\rm app}(\vr,t)$ are respectively the exact and approximated gradients, where the latter is given by the low mobility~\eref{coulowmob} or static~\eref{eq_static_nabla} solutions.
As expected, the accuracy of the first order low mobility expansion becomes worse as $r$ increases and that of the static defect is less sensitive to the distance from the defect center.
However, for distances $r$ comparable to the distance from the annihilation point (which in the representation of Fig.~\ref{error} is set to one) the low mobility expansion clearly leads to an improved approximation as compared to the static solution.

\section{Multi-defect solutions}
 
 \label{sec_multi_defects}
 
So far, our analysis was restricted to solutions of the diffusion equation for a single defect.
As already mentioned, due to the linearity of the diffusion equation~\eref{Diffusioneq} the 
multi-defect solution can be expressed as the linear superposition of single-defect solutions.
Therefore, the counterpart of~\eref{labgrad} for multiple defects with positions $\q_i(t)$, velocities $\vi_i(t)$ and charges $s_i$ is given in the lab frame by
\begin{equation}
    \label{labgrad_mult}
   \nabla\theta(\x,t) = -\sum_i\frac{s_i}{2}\eps \int_{-\infty}^{t}\frac{\rmd t'}{(t-t')} \, \left[\nabla+\vi_i(t')\right]e^{-\frac{|\x-\q_i(t')|^2}{4(t-t')}}.
\end{equation}
In general, most of features of the multi-defect dynamics can be deduced from the single defect solution upon summation. 
However, this is not always the case in particular because the full solution~\eref{labgrad_mult} is nonlocal in time.
In what follows, we illustrate nontrivial consequences of this property by first investigating memory effects in presence of defects pair creation and annihilation and then the orientation mismatch between motile defects.

\subsection{Defect pair creation and annihilation}

 \label{sec_multi-defects-annihilation}

The integral in Eq.~\eref{labgrad_mult} runs from time $-\infty$ as a consequence of the fact that a single defect can never be spontaneously created.
Indeed, because of the topological constraint on the total charge of the system, defects can only be created and destroyed in pairs of opposite charges.
Hence, considering two defects with charges $\pm s$ and positions $\q_{\pm}$ allows to describe a pair creation event at time $\tc$ by imposing that the trajectories $\q_{\pm}(t)$ coincide for all $t < \tc$.
Similarly, the annihilation of the defect pair at time $\ta$
is modeled imposing that $\q_{+}(t) = \q_{-}(t)$ for all $t > \ta$.
The resulting orientation field then satisfies
\begin{eqnarray}
    \label{two_defects_angle}
   \fl \nabla\theta(\x,t) = -\frac{s}{2}\eps \int_{\tc}^{{\rm min}(\ta,t)} & \frac{\rmd t'}{(t-t')} \left\{ \left[\nabla+\vi_+(t')\right]e^{-\frac{|\x-\q_+(t')|^2}{4(t-t')}} 
   - \left[\nabla+\vi_-(t')\right]e^{-\frac{|\x-\q_-(t')|^2}{4(t-t')}}
   \right\}, \nonumber \\
\end{eqnarray}
where it is clear that taking $\q_{+}(t) = \q_{-}(t)$ and $\vi_{+}(t) = \vi_{-}(t)$ the integrand vanishes as required.

Unlike the pseudo-static expansions derived in Sec.~\ref{sec_expansions}, Eq.~\eref{two_defects_angle} grants us access to the relaxation dynamics of the orientation field in response to the local perturbations created by the creation and annihilation of defects.
Therefore, taking $\tc < t < \ta$ and
expanding~\eref{two_defects_angle} for $\x$ far from $\q_{\pm}(t)$, we find that the disturbance generated by the creation of a defect pair scales as
\begin{eqnarray} \label{field_creation}
|\nabla\theta(\x,t)| & \simeq 
\left|\frac{s}{2}\eps\int_{t_c}^{t} \frac{\rmd t'}{(t-t')}
\left[\vi_{+}(t')-\vi_{-}(t')\right] e^{-\frac{x^2}{4(t-t')}}\right| \nonumber \\
& \propto \frac{(t-t_c)}{x^2} \,\exp\left[-\frac{x^2}{4(t-t_c)}\right] \quad (\tc < t < \ta).
\end{eqnarray}
Contrary to the far field expansion~(\ref{farfirst}) which predicts a power law decay of the gradient amplitude with the distance to the defect pair, 
Eq.~\eref{field_creation} shows that on distances above the diffusive scale $\sim (t - t_c)^{1/2}$ this decay is exponential.
For $x \gg (t - t_c)^{1/2}$ the orientation field thus remains essentially unperturbed by the defects, as expected from the diffusive dynamics of $\te(\x,t)$.

We now investigate the relaxation of $\nabla \theta(\x,t)$ after a defect pair annihilation event at time $\ta$. 
To study how the gradient~\eref{two_defects_angle} vanishes at $t \gg \ta$, we assume that in the time interval $[\tc;\ta]$ the defect trajectories are spread over a typical scale $l \simeq |\q_{+}(t')-\q_{-}(t')|$.
The limit $|\x| \gg l$ clearly corresponds to Eq.~\eref{field_creation} such that here we are instead interested in the case where $|\x| \le l$.
Moreover, taking the long time limit $l^2\ll t-\ta$ and expanding the exponentials we get, after integrating by parts in the velocities,
\begin{equation*}
    \nabla\te(\x,t) \simeq \frac{s}{2}\eps \int^{\ta}_{\tc} \frac{\rmd t'}{(t-t')^2} \left[ \q_+(t') - \q_-(t') \right] 
    \qquad (t > \ta),
\end{equation*}
where we have used that the surface term vanishes due to the conditions $\q_+(t_{\rm a,c})=\q_-(t_{\rm a,c})$.
Hence, in the long time limit the scaling of the gradient can be obtained by substituting $\q_{+}(t')-\q_{-}(t')$ with the associated characteristic scale $l \simeq |\q_{+}(t')-\q_{-}(t')|$, such that:
\begin{equation}
\label{texp}
\fl |\nabla\te(\x,t)| \simeq 
   l \int^{\ta}_{\tc} \frac{\rmd t'}{(t-t')^2}=
l\left(\frac{1}{t-t_a}-\frac{1}{t-t_c}\right)\simeq
  \cases{
\frac{l}{t-t_a} & ($t>\ta \gg \tc$)\\
\frac{l (\ta - \tc)}{t^2} & ($t \gg \ta > \tc$)}.
\end{equation}
Thus, the amplitude of the orientation gradient following a defect annihilation event decays algebraically over long times.
Again, this is in contrast with the results provided by the expansions carried out in Sec.~\ref{sec_expansions} which imply an infinitely fast uniformization of the orientation field. 
Remarkably, Eq.~\eref{texp} predicts that the exponent of the algebraic decay of $|\nabla\te|$ varies with the life-time of the defect pair.
In particular, the perturbation following from the presence of long-lived defects ($\ta \gg \tc$) decays more slowly ($\sim t^{-1}$) than that corresponding to a short-lived pair ($\sim t^{-2}$).

Note also that Eq.~\eref{texp} is only valid in the long time regime for which $l^2 \ll t-\ta$. 
The scaling of $\nabla\te$ in the vicinity of the defect pair at intermediate times $t-\ta < l^2 < t-\tc$ is instead quite nontrivial, at one cannot simplify~\eref{two_defects_angle} by approximating the exponential functions.
This case in fact corresponds to the regime where the perturbation due to the creation of the defect pair has diffused over the scale $l$ but the field has not yet relaxed over this scale after the annihilation of the defects at time $\ta$.
In this case the gradient~\eref{two_defects_angle} thus explicitly depends on the details of the defects trajectories.

\subsection{Mismatched rotating defects}

 \label{sec_multi-defects-mismatch}

In Sec.~\ref{sec_far_field} we investigated the far field limit of the orientation field $\theta(\x,t)$ which for a defect moving with a long-time averaged angular momentum $\eL$ is given in the lab reference frame by
\begin{equation}
    \label{lab_farang}
    \theta(\x,t)=s\arg\left[\x - \q(t)\right]+\frac{s \eL}{2}\ln\left[\frac{|\x - \q(t)|}{\lambda}\right]+\te_0 ,
\end{equation}
where as before we included the scale $\lambda$ to nondimensionalize the argument of the logarithm and $\theta_0$ as an integration constant. 
As previously discussed, a defect moving with nonzero angular momentum generates an orientation gradient with a radial contribution,
which in turn gives rise to tangential effective forces between defects.

\begin{figure}[t!]
    \centering
  \includegraphics[width=\linewidth]{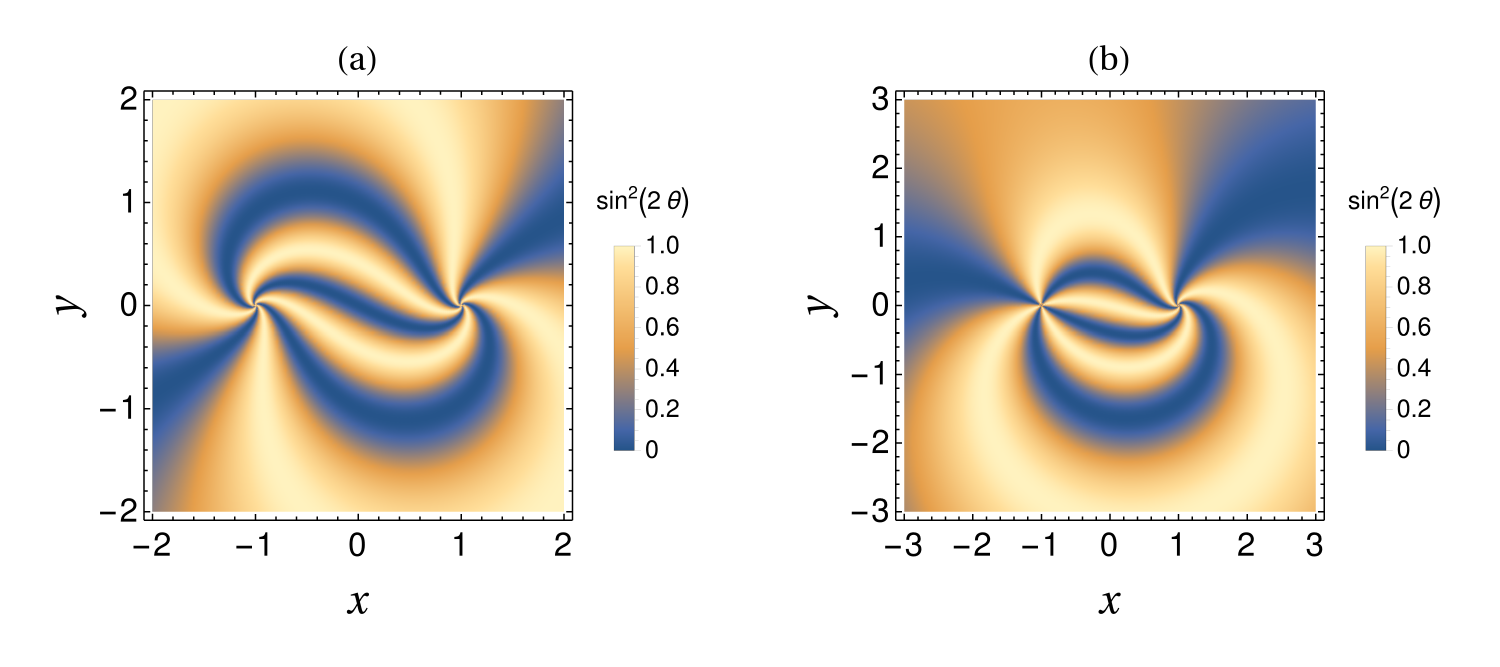}
    \caption{Map of \colrev{$\sin^2 2\te(\x,t)$} for a defect pair with charges $s = \pm 1$ at positions $\q_{\pm} = \pm \hat\x$. 
    Panels (a) and (b) respectively correspond to equal ($\eL_+=\eL_-=1$) and different ($\eL_+=1$, $\eL_-=0$) angular momenta.}
    \label{fig:my_label}
\end{figure}
To study this phenomenon more in depth, we now consider a pair of defects of charges $\pm s$ moving along circular trajectories of radius $R_{\pm} \ll |\q_+(t) - \q_-(t)|$. 
Denoting their angular momenta $\eL_{\pm}$, Eq.~\eref{lab_farang}
generalizes to\\
\begin{eqnarray}
\label{generalmismatch}
\te(\x,t) - \theta_0 & = s\left\{
\arg\left[\x - \q_+(t)\right] - \arg\left[\x - \q_-(t)\right]\right\} \nonumber \\
& + \frac{s}{2}\left\{\eL_{+}\ln\left[\frac{|\x - \q_+(t)|}{\lambda_+}\right]-\eL_{-}\ln\left[\frac{|\x - \q_-(t)|}{\lambda_-}\right]\right\} .
\end{eqnarray}
Eq.~\eref{generalmismatch} describes a pair of so-called \emph{mismatched} defects~\cite{GiomiOrientation,tang2017orientation}, 
i.e. a pair of defects leading to an orientation field different from that predicted by the static solution for $\eL_{\pm} = 0$.
\colrev{Figure~\ref{fig:my_label} shows two examples of such mismatched configurations.}
In fact, taking $\eL_+ = \eL_- = \eL$ we recover, 
upon redefinition of the parameters,
the mismatched solution derived by Tang and Selinger~\cite{tang2017orientation} by imposing a finite mismatching angle $\delta\theta \sim s\eL$ between two defects.
In Eq.~\eref{generalmismatch}, on the contrary, mismatching occurs spontaneously and is a direct consequence of the fact that the orientation field $\te$ keeps the memory of the history of the defects.
 Although the origin of mismatch is more easily understood from the far field approximation Eq.~\eref{generalmismatch}, we emphasize that for a collection of moving defects the full solution~\eref{two_defects_angle} shall in many cases correspond to mismatched configurations.

Interestingly, taking $\eL_+ \ne \eL_-$ in~\eref{generalmismatch}, it is straightforward to show that the associated free energy of the configuration is infinite.
Indeed, for Eq.~\eref{generalmismatch} to be valid infinitely far away from the defect pair the defects must have spun from an infinite time, which requires an infinite amount of energy.
A finite free energy of the system is then naturally recovered considering defects that have started to rotate from a finite time $T$.
Similarly to the defect creation phenomenology discussed above, a similar calculation as the one leading to Eq.~\eref{field_creation} reveals that the the diverging contribution to~\eref{generalmismatch} is negligible on scales $\gg T^{1/2}$,
which ensures that the total free energy of the system remains finite.

\section{Discussion}

\label{sec_discussion}

We have derived the exact two-dimensional orientation field generated by a collection of moving topological defects. 
As we repeatedly emphasized, the main addition of our work to the existing literature is the derivation of the full time dependency of the orientation field in terms of the defect dynamical degrees of freedom.
Indeed, our calculation reveals striking features which are not captured by previous approaches.

 For instance, our results indicate that defects carrying angular momentum generate spiralling force fields beyond the typical scale of their rotating motion (Eq.~\eref{fargrad}). 
This property moreover generally leads to mismatched defect configurations even in absence of elastic anisotropy or imposed boundary conditions on the orientation. 
This indicates that in generic situations defects may spontaneously annihilate following curved trajectories simply because the orientation field has kept the memory of their past history and is not accurately described by the static solution~\eref{eq_static_nabla}.
We moreover stress that the mismatched solution~\eref{generalmismatch} is conceptually different from that derived in Ref.~\cite{tang2017orientation}.
Indeed, Eq.~\eref{generalmismatch} is formally only valid beyond the typical scale of defects motion, such that it remains compatible with the near field expansion result~\eref{nearexp} that takes a different functional dependency in the defect velocity.
The expression derived in Ref.~\cite{tang2017orientation}, on the contrary, is assumed valid down to a cutoff scale given by the defect core radius and thus becomes incompatible with our near field expansion when this cutoff is made arbitrary small.

We furthermore showed in Sec.~\ref{sec_multi-defects-annihilation} that the amplitude of the orientation field gradient decays algebraically in time after the annihilation of a defect pair (Eq.~\eref{texp}). 
Therefore, the orientation field $\te(\x,t)$ may keep a long-time memory of the presence of the defects, even after they have annihilated.

In all this work we have worked on the full $\R^2$ plane assuming uniform boundary conditions at infinity. 
In contrast, it is well known that the presence of boundaries substantially affects the behavior of defects~\cite{Review_LC_Harth2020,Cortese2018PRE}.
Thus, to account more quantitatively for realistic situations, the present work would have to be generalized to support general boundary conditions. 
We do not expect this point to raise major difficulties at it can be straightforwardly addressed using an appropriately modified Green's function.

Further extensions of the formalism studied here to more complex scenarios include the presence of backflow~\cite{TothPRL2002,Sven2003PRL}, elastic anisotropy~\cite{Brugues2008PRL}, activity~\cite{DoostmohammadiActNem,Thampi2014EPL}, or the extension to three dimensions.
These, however, quickly becomes more complex as in all these cases the dynamics of $\te(\x,t)$ becomes nonlinear. 
Nevertheless, in some cases perturbative derivations based on the solution of the linear problem might still be possible.

\colrev{We shall finally comment on the fact throughout this work the trajectory of the defect was assumed to be known. 
Several approaches exist to derive $\q(t)$ from the order parameter dynamics (e.g. Eq.~\eref{RelGL})~\cite{PismenBook,DennistonPRB1996,Radzihovsky2015PRL,TangSM2019,Vafadefects2020,AnghelutaNJP2021}.
How the dynamical contributions to the solution~\eref{labgrad} affect the single and multi-defect dynamics will be addressed in a future work~\cite{nextpaper}.}

\section*{References}
\bibliography{bibliography}

\appendix

\addtocontents{toc}{\addtolength{\cftsecnumwidth}{5em}} 
\addtocontents{toc}{\addtolength{\cftsubsecnumwidth}{4em}} 

\section{Derivation of the orientation gradient for a moving defect}

\label{app_nablathe}

In this section, we present additional calculation details for the derivation of the expression of the orientation field created by a moving defect given in Eq.~\eref{refgrad}.
We start from the expression for $\nabla\varphi(\vr,t)$ given in~\eref{eq_nabla_phi1}:
\begin{eqnarray}
    \nabla\varphi(\vr,t) & = & -\eps\int_{{\cal C}_a}\rmd\bi{l}\int\rmd t' \, G(\vr - \y,t,t') (\vi(t')\cdot \nabla)\thst(\y) \nonumber \\
    & & + \int\rmd^2y\int\rmd t' \, G(\vr - \y,t,t') ( \vi(t')\cdot \nabla)\nabla\thst(\y),
    \label{eq_nabla_phi1app}
\end{eqnarray}
To evaluate the boundary term in~\eref{eq_nabla_phi1app}, we consider a circle ${\cal C}_a$ of radius $a \to 0$ around the defect.
Using the expression of $\nabla\thst$ given in~\eref{eq_static_nabla}, we find after some algebra that the nonvanishing contribution for $a=0$ reads
\begin{eqnarray*}
\fl \eps\int_{{\cal C}_a}\rmd\bi{l} \, G(\vr - \y,t,t') (\vi(t')\cdot \nabla)\thst(\y) & =
- s\int_0^{2\pi}\rmd\vartheta\, \hat{\bi{e}}(\vartheta)G(\vr,t,t') \left[\vi(t') \cdot \eps \hat{\bi{e}}(\vartheta)\right] + {\cal O}(a),
\\
& \underset{a \to 0}{=} \pi s \eps \, \vi(t') G(\vr,t,t') ,
\end{eqnarray*}
where $\hat{\bi{e}}(\vartheta)$ is the unit vector oriented along the direction set by $\vartheta$.
Integrating by parts the second term on the r.h.s.\ of~\eref{eq_nabla_phi1app} we get another boundary contribution
\begin{eqnarray*}
   \fl \eps\int_{{\cal C}_a}\rmd\bi{l} \cdot \vi(t') \, G(\vr - \y,t,t') \nabla\thst(\y) & =
- s\int_0^{2\pi}\rmd\vartheta\, \vi(t') \cdot \hat{\bi{e}}(\vartheta)G(\vr,t,t') \eps \hat{\bi{e}}(\vartheta) + {\cal O}(a),
\\
& \underset{a \to 0}{=} -\pi s \eps \, \vi(t') G(\vr,t,t') ,
\end{eqnarray*}
such that the resulting expression for $\nabla\varphi(\vr,t)$ reads
\begin{eqnarray}
    \nabla\varphi(\vr,t) & = & -2\pi s \eps  \int\rmd t' \, \vi(t') G(\vr,t,t') \nonumber \\
    & & + \int\rmd^2y\int\rmd t' \, ( \vi(t')\cdot \nabla_{\vr})G(\vr - \y,t,t') \nabla\thst(\y).
    \label{eq_nabla_phi2app}
\end{eqnarray}

As discussed in the main text, using the relation satisfied by the Green's functions
\begin{equation*}
    (\vi(t')\cdot\nabla)G(\vr,t,t') = (\partial_{t'} + \nabla^2)\left[G_{\rm D}(\vr,t-t') - G(\vr,t,t')\right],
\end{equation*}
with $G_{\rm D}(\vr,t-t')$ the Green's function of the diffusion equation,
we recast Eq.~\eref{eq_nabla_phi2app} into
\begin{eqnarray}
    \fl & \nabla\varphi(\vr,t) = & -2\pi s \eps  \int\rmd t' \, \vi(t') G(\vr,t,t') \nonumber \\
    \fl & & + \int\rmd^2y\int\rmd t' \, (\partial_{t'} + \nabla^2) \left[G_{\rm D}(\vr - \y,t-t') - G(\vr - \y,t,t')\right] \nabla\thst(\y).
    \label{eq_nabla_phi3app}
\end{eqnarray}
Clearly, the $\sim \partial_{t'}$ vanish, while we calculate the $\sim \nabla^2$ terms by integrating again by parts.
Namely, considering a generic function $\Gamma(\vr,t,t')$, we get after integrating by parts twice
\begin{eqnarray} \label{eq_Gamma_IBP}
    \fl \int\rmd^2y \,\nabla^2 \Gamma(\vr - \y,t,t')\nabla\thst(\y) = & \int_{{\cal C}_a}\left[(\eps\rmd\bi{l})\cdot \nabla_{\y} \Gamma(\vr-\y,t,t')\right] \nabla\thst(\y) \nonumber \\
    & - \int_{{\cal C}_a}\left[(\eps\rmd\bi{l})\cdot\nabla\nabla\thst(\y)\right]  \Gamma(\vr - \y,t,t') \nonumber \\
    & + \int\rmd^2y \, \Gamma(\vr - \y,t,t') \nabla \nabla^2\thst(\y).
\end{eqnarray}
The last term on the r.h.s.\ of~\eref{eq_Gamma_IBP} vanishes due to the condition $\nabla^2\thst = 0$.
The first term can be treated similarly to the boundary term in~\eref{eq_nabla_phi1app}, which leads to 
\begin{equation*}
    \int_{{\cal C}_a}\left[(\eps\rmd\bi{l})\cdot \nabla_{\y} \Gamma(\vr-\y,t,t')\right] \nabla\thst(\y) = \pi s \eps \nabla \Gamma(\vr,t,t').
\end{equation*}
As the second term involves second order derivatives of $\thst$, we expand $\Gamma$ in the integrand w.r.t.\ $\y$ in order to get
\begin{eqnarray*}
    \fl -\int_{{\cal C}_a}[(\eps\rmd\bi{l})\cdot\nabla\nabla\thst(\y)]  \Gamma(\vr - \y,t,t') & = -\frac{s}{a}\int_0^{2\pi}\rmd\vartheta
    \eps \hat{\bi{e}}(\vartheta)
    \left(1 - a \hat{\bi{e}}(\vartheta) \cdot\nabla \right)\Gamma(\vr,t,t') + {\cal O}(a) \\
    & \underset{a\to 0}{=} \pi s \eps \nabla \Gamma(\vr,t,t'),
\end{eqnarray*}
where we have used that the ${\cal O}(a^{-1})$ contribution vanishes by symmetry.
Replacing $\Gamma = G_{\rm D} - G$, we thus obtain from~\eref{eq_nabla_phi3app} the expression of $\nabla\varphi(\vr,t)$ given in~\eref{eq_nabla_phi_final}.

As presented in the main text, using the relation between $\nabla\thst$ and $G_{\rm D}$ the resulting expression for the full solution $\nabla\theta$ reads
\begin{equation}
\label{refgrad_app}
    \nabla\theta(\vr,t) = -2\pi s \eps \int_{-\infty}^{+\infty}\, \rmd t' \left[ \nabla + \vi(t') \right] G(\vr,t,t') .
\end{equation}
Because $\theta$ is a multivalued function, the r.h.s.\ of~\eref{refgrad_app} is not written as the gradient of a scalar function.
Nevertheless, we show now that $\nabla\theta(\vr,t)$ is irrotational almost everywhere on the plane.
Namely, we find
\begin{eqnarray}
    \fl (\eps\nabla) \cdot \nabla\theta(\vr,t) & = 
    -2\pi s \int_{-\infty}^{+\infty}\, \rmd t' \, \left[ \nabla^2 + \vi(t')\cdot\nabla \right]G(\vr,t,t') \nonumber \\
    \fl
    &=-2\pi s \int_{-\infty}^{+\infty} \,\rmd t'\,\left[ \left(\partial_t +(\vi(t')-\vi(t))\cdt\nabla\right) G(\vr,t,t')-\delta^2(\vr)\delta(t-t')\right],\nonumber
\end{eqnarray}
where to obtain the second line we have used the definition of $G$.
Finally, using the identity $(\partial_t - \vi(t)\cdot\nabla)G(\vr,t,t')
= -(\partial_{t'} + \vi(t')\cdot\nabla)G(\vr,t,t')$ leads to
\begin{equation}
    (\eps\nabla) \cdot \nabla\theta(\vr,t) = 
    \left[\nabla \times \nabla\theta(\vr,t)\right]\cdot\hat{\z} = 2 \pi s\delta^2(\vr).
\end{equation}

\section{The far field expansion of Eq.~\eref{refgrad}}

\label{app_far}

Here, we detail the calculation steps leading to the far field approximation~\eref{farfirst}.
We keep the same notations as in the main text, in particular with $\ell$ denoting the typical length scale of the defect motion.
Let us split the integral on the rhs of Eq.~\eref{refgrad} into two contributions, one given by the gradient and the second by the defect velocity $\vi(t')$.
Expanding the gradient part of the integral up to leading order in $\ell/r$ yields
\begin{equation*}
\label{farnabla} \fl
    \int_{-\infty}^{t} \frac{\rmd t'}{(t-t')}\nabla\left[ e^{-\frac{|\vr+\Dq|^2}{4(t-t')}}\right]=
    -\frac{\vr}{2}\int_{-\infty}^{t} \frac{\rmd t'}{(t-t')^2} e^{-\frac{r^2}{4(t-t')}}+{\cal O}\left(\frac{\ell}{r}\right)=-2\frac{\hat \vr}{r}+{\cal O}\left(\frac{\ell}{r}\right).
\end{equation*}

For the velocity contribution, we first integrate by parts to obtain
\begin{eqnarray}
    \label{farvelstart} \fl
    \int_{-\infty}^{t}\rmd t' \, \frac{\vi(t')}{(t-t')} e^{-\frac{(\vr+\Dq)^2}{4(t-t')}} = 
    \int_{-\infty}^{t} \rmd t' & \, \frac{\Delta\q(t,t')}{(t-t')^2} e^{-\frac{|\vr +\Delta\q(t,t')|^2}{4(t-t')}} \times \\
    & \left[
    1-\frac{|\vr +\Delta\q(t,t')|^2}{4(t-t')}+\frac{\vi(t')\cdot(\vr +\Delta\q(t,t'))}{2}\right]\nonumber.
\end{eqnarray}
Due to the $\Delta\q(t,t')$ prefactor in the integral, 
at first order in $\ell/r$, we can approximate in~\eref{farvelstart} $\vr +\Delta\q(t,t')$ by $\vr$. 
Then, using the change of variable $u=r^2/(4(t-t'))$ we obtain
\begin{equation}
\label{farvelstart_u}
     \fl
    \int_{-\infty}^{t}\rmd t' \, \frac{\vi(t')}{(t-t')} e^{-\frac{(\vr+\Dq)^2}{4(t-t')}} \simeq 
    \frac{4}{r^2} \int_{0}^{+\infty} \rmd u \,
    \Delta\q(t,u)e^{-u} \left( 1 - u + \frac{\vi(u)\cdot\vr}{2} \right) .
\end{equation}
It is thus easy to see that for large $r$ the dominant term in the r.h.s.\ of~\eref{farvelstart_u} is the one proportional to $\vi\cdot \vr$.

Coming back to the $t'$ variable, we can therefore write that the velocity part of the integral is given up to order $\ell/r$ in index notations by
\begin{equation}
\label{farvelcont} \fl
    \int_{-\infty}^{t} \rmd t' \frac{v_j(t')}{(t-t')}e^{-\frac{(r_k+\Delta q_k(t,t'))^2}{4(t-t')}}=
    \frac{r_i}{2}\int_{-\infty}^{t}\rmd t' \,
    \frac{v_i(t')\Delta q_j(t,t')}{(t-t')^2}e^{-\frac{r_k^2}{4(t-t')}} + {\cal O}\left(\frac{\ell}{r}\right),
\end{equation}
and where summation over repeated indices is implied.
Integrating again by parts the l.h.s.\ of \eref{farvelcont}, it is straightforward to show that for all $i$ and $j$
\begin{equation*}
    \int_{-\infty}^{t}\rmd t' \,
  \frac{v_i(t')\Delta q_j(t,t')+ v_j(t')\Delta q_i(t,t')}{(t-t')^2}e^{-\frac{r^2}{4(t-t')}} 
  = {\cal O}\left(\frac{\ell}{r}\right).
\end{equation*}
Hence, we can anti-symmetrize the integrand on the r.h.s.\ of Eq.~\eref{farvelcont}, which naturally involves the angular momentum of the defect $L(t') = v_1(t')\Delta q_2(t,t')-v_2(t')\Delta q_1(t,t')$:
\begin{equation*}
\label{antsymm}
    \frac{1}{2}[(v_i(t')\Delta q_j(t,t')-v_j(t')\Delta q_i(t,t')]=-\frac{\epsilon_{ij}}{2}L(t') .
\end{equation*}
Going back to vector notations, we thus obtain
\begin{equation*}
     \int_{-\infty}^{t}\rmd t'\,\frac{\vi(t')}{(t-t')}e^{-\frac{(\vr+\Dq)^2}{4(t-t')}}
     = \eps\vr\int_{-\infty}^{t}\rmd t'\frac{L(t')}{4(t-t')^2}e^{-\frac{\vr^2}{4(t-t')}}+{\cal O}\left(\frac{\ell}{r}\right).
\end{equation*}
Multiplying the integral by $- s\eps /2$ and combining everything together finally leads to Eq.~\eref{farfirst} of the main text. 

\section{The low mobility expansion for defects interacting via the Coulomb force}

\label{app_lowmob-xi}

The expression of the gradient~\eref{refgrad} evaluated at the position $\vr = -\q(t)$ reads
\begin{equation*}
  \fl \qquad \nabla\te(-\q(t),t) = -\frac{s}{2}\eps\int_{-\infty}^{t} \frac{\rmd t'}{(t-t')} \left(\frac{\q(t')}{2(t-t')}+\vi(t')\right)e^{-\frac{q^2(t')}{4(t-t')}}.
\end{equation*}
Below we treat the $\propto \q(t')$ and $\propto \vi(t')$ parts of the integrand separately.
Using $\q(t')=\sqrt{q^2(t)+2\mu (t-t')} \, \uveci$ and substituting $u= q(t)^2/[4(t-t')]$, we obtain
\begin{equation}
\label{integratedq}
    \int_{-\infty}^{t}\rmd t' \frac{\q(t')}{2(t-t')^2}e^{-\frac{q^2(t')}{4(t-t')}}= \frac{4\sqrt{\pi}}{q(t)} 
    e^{-\frac{\mu}{2}}U\left(-\frac{1}{2},0,\frac{\mu}{2}\right)\uveci ,
\end{equation}
where $U$ denotes the confluent hyper-geometric function, which for ${\rm Re}(a)>0$ is defined as:
\begin{equation*}
    U(a,b,z) = \frac{1}{\Gamma(a)}\int_0^\infty \rmd x \,
    e^{-z x} x^{a-1} (1+x)^{b-a-1},
\end{equation*}
while the definition can be analytically continued to $a=-\frac{1}{2}$ by using the relation $U(a,b,z)=z^{1-b}U(1+a-b,2-b,z)$.
Expanding \eref{integratedq} up to order $\mu$, we find
\begin{equation}
\label{integratedq_exp}
    \frac{4\sqrt{\pi}}{q(t)} 
    e^{-\frac{\mu}{2}}U\left(-\frac{1}{2},0,\frac{\mu}{2}\right)\uveci
    \simeq
    \frac{2}{q(t)} \left[1-\frac{\mu}{2}+\frac{\mu}{4}\ln\left(\frac{8 e^{1-\gamma_{\rm E}}}{\mu}\right)\right] \uveci .
\end{equation}
We moreover derive the second part of the integral in an analogous way, which yields
\begin{equation}
    \label{integratedv}
    \int_{-\infty}^{t} \rmd t' \frac{\vi(t')}{(t-t')}e^{-\frac{q^2(t')}{4(t-t')}}
    \simeq
    -\frac{\mu}{q(t)}\ln\left(\frac{8e^{-\gamma_{\rm E}}}{\mu}\right) \uveci .
\end{equation}
Combining \eref{integratedq_exp} and \eref{integratedv} we finally recover~\eref{specvalue} of the main text.

\end{document}